\documentclass[epj]{svjour}

\usepackage{t1enc}
\usepackage[latin2]{inputenc}
\usepackage{color}
\usepackage{caption,subfig}
\usepackage{hyperref}

\usepackage{savesym}
\usepackage{txfonts, nicefrac}
\savesymbol{iint}
\savesymbol{iiint}
\savesymbol{iiiint}
\savesymbol{idotsint}
\usepackage{amsmath,amsfonts,amssymb}
\savesymbol{degree}
\savesymbol{vec}
\restoresymbol{TXF}{iint}
\restoresymbol{TXF}{iiint}
\restoresymbol{TXF}{iiiint}
\restoresymbol{TXF}{idotsint}
\restoresymbol{STD}{degree}
\usepackage{MnSymbol}
\usepackage{bbm}

\usepackage{graphics}
\usepackage{float}
\usepackage{epsfig}

\newcommand{\ve}[1]{\mathbf{#1}}

\newcommand{\avr}[1]{\langle #1\rangle}
\definecolor{!R}{rgb}{1,0,0}

\begin{document}
\title{Multicomponent Modified Boltzmann Equation and Thermalization}
\author{M. Horv\'ath\inst{1}\inst{2}\and T. S. Bir\'o\inst{2}}
\institute{Department of Theoretical Physics, Wigner Research Centre for Physics, Institute for Particle and Nuclear Physics, H-1525 Budapest, Konkoly Thege Mikl\'os \'ut 29-33, Hungary
\and Department of Theoretical Physics, Budapest University of Technology and Economics, Budafoki \'ut 8. H-1111 Budapest, Hungary, \email{horvath.miklos@wigner.mta.hu}}

\abstract{
The existence of stationary distributions in a multicomponent Boltzmann equation using a non-additive kinetic energy composition rule for binary collisions is discussed. It is found that detailed balance is not achieved when -- in contrast to the case of a single rule -- several different composition rules are considered. The long-time behaviour of a simple momentum space model is explored numerically: saturating, heating and cooling solutions are presented. These results may be used in modelling the kinetics of multicomponent systems, such as hadronic fireballs or quark-gluon plasma.
\PACS{{05.20.Dd}{} \and {05.70.Ln}{} \and {52.27.Cm}{}}
}
\maketitle

\section{Introduction}
\indent Kinetic theory with the Boltzmann equation (BE) in its core is an efficient and widely used approach to describe weakly interacting quasi-particle systems, like rare gases. In these systems the basic interaction events are instantaneous, binary collisions. In a homogeneous approximation the state of the system at a given time instant is characterized by the particle momenta. The BE governs the time evolution of the phase-space density, $f$, by summing up probabilities of events wherein a particle with a given momentum is scattered in or out of a small volume element of the phase-space in a unit time. Assuming that the particles forget their history between two consecutive collisions, the product of two density functions enters into the binary collision integral:\\ \indent
\begin{equation}\label{BE}
\frac{\partial}{\partial t}f_1=\int_{234}\delta^{(4)}(\textnormal{en.-mom. conserv.})w_{1234}(f_3f_4-f_1f_2)=:\mathcal{I}_1.
\end{equation}
The lower indices refer to phase-space coordinates in this shorthand notation (e.g. the momentum of the particle in the case of homogeneous gas). \\ \indent
There are many possible ways to elevate the strict restrictions of elastic collision and to investigate more complex dynamics, keeping on the concept of binary collisions as elementary events. The need for modification emerges when the quasi-particles are not point-like, or long-range interactions arise \cite{coulomb,spicka1,spicka2,spicka3,knoll2}. \\ \indent
Let us regard two possible modifications of the original form of the BE. One can $i)$ use other constraints instead of the ones of elastic collision, or $ii)$ abandon the product structure of the collision kernel. In the first case the composition rule for conserved quantities are modified, while in the second case two-particle correlations are taken into account in a certain way. Several examples have been discussed in the literature \cite{biro3,kaniadakis,biro4,kaniadakis2,lima}. In this paper we exploit the modification of the energy composition rule in details, in particular its possible extension to multicomponent systems. \\ \indent 
In general, the assumption of instantaneous pairwise collisions is getting worse as the interaction becomes stronger. In the kinetic theory one aims to describe the system on a longer time scale compared to the characteristic time period of a single collision. There are several examples for deriving the kinetic description from a microscopic theory, the most famous one is by Kadanoff and Baym \cite{kadanoff_baym}. Other examples can be found in the literature as well \cite{greiner1,cassing1,knoll1,spicka3}. These approaches distinguish between microscopic and mesoscopic time scales. However, it is not guaranteed that all consistent approximation schemes lead to the BE. Let us mention a few examples wherein the microscopic details modify the original picture. In \cite{spicka1} the authors emphasize the role of non-instantaneous collision processes. They argue that the time and space arguments of the functions in the collision term of the BE differ because of non-local corrections. Other examples \cite{coulomb,spicka2,spicka3,knoll2,buss} (e.g. dense plasma of electrons, the Fermi-liquid in semiconductors or the kinetic description of high-energy nuclear collisions) show the increasing importance of off-mass-shell scattering processes in dense systems. \\ \indent
We present now a heuristic argument how the modification of energy composition can emerge. In the ordinary BE, all the particles are on mass shell $p^0=E(\ve{p})$. Consider one of the colliding particles off-mass-shell with a spectral density
\begin{equation}
\rho_\gamma(p^0-E_{\ve{p}}) = \frac{1}{\pi}\frac{\gamma}{(p^0-E_{\ve{p}})^2+\gamma^2} \xrightarrow{\gamma \rightarrow 0} \delta(p^0-E_{\ve{p}}).
\end{equation}
We suppose that $\gamma$ could take on several different values, stochastically collision-by-collision. Let us assume moreover that the random variable $\gamma$ has small variance and its expectation value is large, compared to the typical energy scale of the collisions. Now, we perform an averaging over $\gamma$, similarly to the reasoning appeared in \cite{rafelski}. The resulting $\mathcal{I}=\llangle \mathcal{I}_\gamma \rrangle:=\int_0^\infty\mathrm{d}\gamma g(\gamma)\mathcal{I}_\gamma$ is the mesoscopic collision integral. To analyse the effect of the averaging, we write the r.h.s. of the BE (\ref{BE}) into the following compact form: $~\mathcal{I}_\gamma =\int_{234}\!\rho_\gamma(E_1+E_2-E_3-E_4)\mathcal{K}_{1234} $, where $~\mathcal{K}_{1234}=w_{1234}\delta^{(3)}({\sum_i}'\ve{p}_i)(f_3f_4-f_1f_2)$, and ${\sum_i}'x_i:=x_1+x_2-x_3-x_4$. It is apparent that $\rho_\gamma$ takes over the role of the mass shell constraint on the kinetic energy. Assuming that the integration by $\gamma$ and by the phase-space variables can be interchanged, we arrive at
\begin{eqnarray} \label{averaging}
\mathcal{I} &=& \Big\llangle\int_{234}\!\! \mathcal{K}_{1234}\rho_{\gamma,2}\Big\rrangle =\int_{234}\!\! \mathcal{K}_{1234}\llangle \rho_\gamma\rrangle_2 \approx \int_{234}\!\!\left(\left.\mathcal{K}_{1234}\right|_{E_2=\Omega+\Delta} +\dots\right)\llangle \rho_\gamma\rrangle_2 \approx \int_{234}\delta(E_1+E_2-E_3-E_4-\Delta)\mathcal{K}_{1234},
\end{eqnarray}
where $\Delta$ refers to the position of the peak of the smeared spectral density $\llangle\rho_\gamma\rrangle$ and $\Omega=E_3+E_4-E_1$ (for a detailed derivation see Appendix \ref{app1}). \\ \indent
We do not specify the mechanism encoded in the presence of the noise on $\gamma$. Some physics arguments however can be mentioned. $i)$ In a medium particles can have a thermal mass due to the heat bath \cite{biro5}. $ii)$ An external force stemming from a momentum-dependent, so called optical potential, is not taken into account by a Vlasov-type contribution. Such momentum-space inhomogeneities can be observed in early-time non-Abelian plasmas due to the plasma instabilities \cite{strickland,ipp}. When $iii)$ long-range interaction is present, the propagation of the quasi-particles could be disturbed by the long-wavelength modes of the system. As a consequence, the originally independent two-particle collisions start to "communicate" with each other. The effect of these, beyond-two-particle processes can be incorporated into a mean-field description \cite{biro6,arnold}. In cases when the leading order is still the two-quasi-particle collision, one may deal with these events on the level of a kinetic description, but with modified kinetic energy addition rule. \\ \indent
In the spirit of the argumentation given above, we consider a spectral density of the incoming particles showing two peaks and we get:
\begin{equation}\label{kinequ0}
\mathcal{I} \approx \int_{234}\!\! \left(\delta({\sum_i}' E_i-\Delta^A)+ \delta({\sum_i}' E_i-\Delta^B) \right)\mathcal{K}_{1234}.
\end{equation}
The two distinct constraints suggest a multicomponent treatment, where the collisions between different particle species have different modifications for the respective kinetic energy sums.

\section{Multicomponent kinetic equations with modified constraints}
\indent Let $f^\alpha$ be a one-particle density function of the $\alpha$-kind particles, depending on the momentum-space coordinates and time. It is normalized to unity $\int_1f^\alpha_1=1$ at any time, and satisfies the following kinetic equation:
\begin{equation} \label{kinequ}
\frac{\partial}{\partial t} f^\alpha_1 = \sum\limits_\beta\int_{234}\mathcal{W}_{1234}^{\alpha\beta}(f_3^\alpha f_4^\beta - f_1^\alpha f_2^\beta) =\sum\limits_\beta \mathcal{I}^{\alpha\beta}_1.
\end{equation}
We call (\ref{kinequ}) the multicomponent modified BE (MMBE). The dynamics can be interpreted as a sequence of $~\{(E^\alpha_1,\ve{p}_1),(E^\beta_2,\ve{p}_2)\} \rightarrow \{(E^\alpha_3,\ve{p}_3),(E^\beta_4,\ve{p}_4)\}$ collisions. The number of particles is conserved for all species separately. Therefore changes of species like $\alpha\beta\leftrightarrow\beta\beta$, $\alpha\alpha\leftrightarrow\alpha\beta$ are not allowed in this model. \\ \indent
The probability rate that such an event happens is given by the $\mathcal{W}_{1234}^{\alpha\beta}$ part of the kernel, including now the modified kinetic constraints. $\mathcal{W}$ has the following symmetries in its indices simultaneously: $i)$ the interchangeability of incoming (1,2) and outgoing (3,4) collision partners: $1234 \leftrightarrow 2134$, $\alpha\leftrightarrow\beta$ and $ii)$ microscopic time-reversibility: $1234 \leftrightarrow 3412$. $\mathcal{W}$ can be factorized into a rate part and the $\delta$-s prescribing the constraints of pairwise collisions: $\mathcal{W}^{\alpha\beta}_{1234}=w_{1234}^{\alpha\beta}\delta^{(4)}(\mathrm{constraints})$. The invariants of a binary collision are the total momentum ($\ve{p}_1+\ve{p}_2=\ve{p}_3+\ve{p}_4$, 3 constraints) and a quantity depending on the kinetic energies of the particles on the same side of the process: $E_1^\alpha\oplus^{\alpha\beta}E_2^\beta=E_3^\alpha\oplus^{\alpha\beta}E_4^\beta~$ (1 constraint). The energy composition rule is represented here by the symbol $\oplus^{\alpha\beta}$. It is associative and commutative due to the symmetries mentioned above. It can be approximated by the simple addition up to first order in its variables: $E\oplus^{\alpha\beta}E' \approx E+E'+\mathcal{O}^{\alpha\beta}(E^2,(E')^2,EE')$. We note, that the requirement of associativity reflects the mesoscopic nature of our description. Detailed explanation can be found in Ref. \cite{biro_comp_rule}. \\ \indent
It is possible to rewrite the energy composition rule in an additive form, using the function ${L^{\alpha\beta}(E_i^\alpha\oplus^{\alpha\beta}E_j^\beta)=L^{\alpha\beta}(E_i^\alpha)+L^{\alpha\beta}(E_j^\beta)~}$, which maps the composition of the energies into the sum of single energy-dependent quantities, the so-called \textit{quasi-energies} \cite{biro2}. Such a function can be constructed for every associative composition rules: 
\begin{equation} \label{formlog}
L^{\alpha\beta}(x)=\int_0^x\frac{\mathrm{d}z}{\left.\partial_y (z\oplus^{\alpha\beta}y)\right|_{y=0}}.
\end{equation} 
\textit{Therefore one might view such a binary collision like a usual one, but between particles with modified dispersion relation. However, in this case the dispersion relation depends not only on the particle itself, but also on the collision partner.} \\ \indent
It is an irregular feature of the MMBE that the existence of detailed balance cannot be guaranteed in the general multicomponent case. $\partial_t f^\alpha \equiv 0$ should be fulfilled for every $\alpha$ in such a way that all the kernels of the r.h.s. collision integrals in (\ref{kinequ}) are zero, i.e. $f^\alpha_3 f^\beta_4 - f^\alpha_1f^\beta_2 =0$, for every pair of $\alpha$, $\beta$ and for every value of the phase-space variables 1, 2, 3 and 4 which are satisfying the constraints. We shall write the kinetic equation in a form more convenient for the analysis of the detailed balance state. Thinking of a given two-particle collision, we deal with a 6-dimensional phase-space (three dimension for each particles) which is however constrained by $i)$ the momentum conservation (three restrictions) and $ii)$ the suitable quasi-energy conservation depending on the type of the collision. We have two free parameters left, which means the phase-space is restricted to a surface in each collisions: ${~\mathcal{C}^{\alpha\beta}(\ve{P},K):=\left\{(\ve{p},\ve{p}') \,\,:\,\, \ve{p}+\ve{p}'=\ve{P},\,\, E(p)\oplus^{\alpha\beta}E(p')=K \right\}}$.
After the four constraints are integrated out, a 5-fold integral remains instead of the 9-fold one in (\ref{kinequ}). The kinetic equation takes the form
\begin{equation} \label{kinequ2}
\frac{\partial}{\partial t} f^\alpha_1 = \sum\limits_\beta \int\!\!\mathrm{d}^3\ve{P}\iint\limits_{\mathcal{C}^{\alpha\beta}}\!\!\mathrm{d}^2\sigma^{\alpha\beta} w^{\alpha\beta}_{1234}\left.(f_3^\alpha f_4^\beta -f_1^\alpha f_2^\beta)\right|_{(\ve{p}_1,\ve{p}_2),\,(\ve{p}_3,\ve{p}_4) \in \mathcal{C}^{\alpha\beta}(\ve{p}_1+\ve{p}_2=\ve{P}=\ve{p}_3+\ve{p}_4,E_1\oplus^{\alpha\beta}E_2)}.
\end{equation}
Detailed balance requires that the appropriate kernel vanishes on the constraint surface $\mathcal{C}^{\alpha\beta}$. In order to achieve this, the following equations have to hold for every $\alpha$ and $\beta$ pairs simultaneously:
\begin{eqnarray} \label{detbal}
\left. f^\alpha(E(p))f^\beta(E(p'))\right|_{(\ve{p},\ve{p}')\in\mathcal{C}^{\alpha\beta}(\ve{p}+\ve{p}',E\oplus^{\alpha\beta}E')} &=& \mathcal{M}^{\alpha\beta},
\end{eqnarray}
with $\mathcal{M}^{\alpha\beta}$ independent of $\ve{p}$, $\ve{p}'$. Looking for isotropic equilibrium solution with the ansatz 
\begin{equation}\label{detbal_ansatz}
f^\alpha(E) \sim e^{-\gamma^\alpha L^{\alpha\alpha}(E)},
\end{equation} 
the conditions (\ref{detbal}) with $\alpha=\beta$ are automatically satisfied. This is the familiar detailed balance solution of the single-component modified BE \cite{biro2}. In our case, however, conditions with different $\alpha$ and $\beta$ must be satisfied, too. \\ \indent
Let us investigate the case of two components, $\alpha\in\{A,B\}$. One has to deal with three kind of collisions then: $AA$, $BB$ and $AB=BA$, since all the constraints are commutative. The structure of the MMBE (\ref{kinequ2}) in this case reduces to:
\begin{equation}\begin{array}{cccc}
\frac{\partial}{\partial t} f^A_1 = \mathcal{I}^{AA}_1+\mathcal{I}^{AB}_1, & & &
\frac{\partial}{\partial t} f^B_1 = \mathcal{I}^{BA}_1+\mathcal{I}^{BB}_1. 
\end{array}\end{equation}
In the state characterized by (\ref{detbal_ansatz}), all the $\mathcal{I}^{\alpha\alpha} \equiv 0$ by construction of the density functions. The problem is, that supposing the detailed balance of a given kind of collision ($AA$, $BB$, $\mathcal{I}^{AA}\equiv 0$, $\mathcal{I}^{BB}\equiv 0$), the rest ($AB$) will not be fulfilled with the same type of density functions ($\mathcal{I}^{AB}\neq 0$). (There is, however, a trivial solution, namely when all the $L^{\alpha\beta}$ functions are the same.) \\ \indent
Let the collisions between the particles in the same type be in detailed balance ($\mathcal{I}^{AA}\equiv 0$, $\mathcal{I}^{BB}\equiv 0$). For the sake of simplicity, let us suppose $L^{AA} \equiv L^{BB}$. In this case $f^A_1 \sim f^B_1 \sim e^{-L^{AA}_1/T}$, with a phase-space independent $T^{-1}:=\gamma^A=\gamma^B$. Then the kernel of the mixed collisional term $\mathcal{I}^{AB}$ is proportional to
\begin{equation}
\mathrm{kernel\,\,of\,\,} \mathcal{I}^{AB} \sim e^{-\frac{1}{T}L^{AA}_3-\frac{1}{T}L^{AA}_4}-e^{-\frac{1}{T}L^{AA}_1-\frac{1}{T}L^{AA}_2}.
\end{equation}
In the detailed balance state it vanishes for all phase-space points lying on $\mathcal{C}^{AB}$. With no further assumptions for $L^{\alpha\beta}$, the achievement of such a state implies the relation $\mathcal{C}^{AA} \subseteq \mathcal{C}^{AB}$. Starting on the other hand from $\mathcal{I}^{AB} \equiv 0$, the above reasoning leads to $\mathcal{C}^{AB} \subseteq \mathcal{C}^{AA}$. In conclusion, if one does not have other conditions for the modification except the symmetry properties mentioned above, the only detailed balance solution is $\mathcal{C}^{AA}=\mathcal{C}^{AB}$, that is when all the modifications are the same. This is fulfilled only for the one component matter.\\ \indent 
Our result does not imply that the system can not saturate to a time-independent state: $f^{A,B}(p,t) \xrightarrow{t\rightarrow\infty} f^{A,B}_\mathrm{Eq.}(p)$. According to the previous argument, when the modification is coupled to the dynamics in such a way that $~\mathcal{C}^{AB} \xrightarrow{t\rightarrow\infty} \mathcal{C}^{AA}$, saturation behaviour may arise. But, since the convergence in that sense is not restrictive enough, it does not guarantee the asymptotic equilibration. \\ \indent
We note here that some authors emphasized the non-universal nature of the modification of the energy addition, in the sense it should depend on dynamical details of the system \cite{wang}.

\section{Numerical results for long-time behaviour}
\subsection{The model}
\indent In this section we investigate the time evolution of a two-component MMBE (\ref{kinequ}). For this purpose we use a simple toy model. From here on we consider only the isotropic case when all functions $f^\alpha$ depend on the phase-space position through the single-particle energy only. If the collision does not prefer any direction for some reason, or with other words $w_{1234}^{\alpha\beta}$ depends on $|\ve{p}_1-\ve{p}_2|$ and $|\ve{p}_3-\ve{p}_4|$ only, it is reasonable to expect an isotropic state after appropriately long evolution, irrespective to the initial state. If external fields are not presenting, isotropisation is usually much faster than equilibration. \\ \indent
Three elements of the model have to be fixed: $i)$ the energy addition rule in the constraint, $ii)$ the properties of the particles building up the ensemble and $iii)$ the rate function $w^{\alpha\beta}_{1234}$ incorporating the dynamical details of the collisions. Let us specify the last two first: we consider Newtonian particles with dispersion relation $E(p)=\frac{1}{2m}p^2$. We choose the rate function in a way that the system can reach every outgoing state which fulfil the kinematics (i.e. the modified constraint for the energy- and momentum-conservation) with equal probability.\\ \indent
At this point the MMBE (\ref{kinequ2}) reads:
\begin{equation}
\frac{\partial}{\partial t} f^\alpha(p) = \sum_\beta\int\!\!\mathrm{d}^3\ve{p}'\!\!\int\!\!\mathrm{d}^3\ve{q} \frac{1}{\mathcal{Z}^{\alpha\beta}}\delta(E(p)\oplus^{\alpha\beta}E(p') -E(q)\oplus^{\alpha\beta}E(q^*))\left\{f^\alpha(q)f^\beta(q^*)-f^\alpha(p)f^\beta(p')\right\}, \label{kinequ3}
\end{equation}
\begin{equation}
\mathcal{Z}^{\alpha\beta}(P,K^{\alpha\beta},x) = \int\!\!\mathrm{d}^3\ve{q}\delta(K^{\alpha\beta} -E(q)\oplus^{\alpha\beta}E(|\ve{P}-\ve{q}|)) = \frac{2\pi}{P}\int\limits_0^\infty\!\!\mathrm{d}q \frac{q\,\Theta(|y^*|\leq 1)Q^{\alpha\beta}(q)}{g^{\alpha\beta}( E(q),E(Q^{\alpha\beta})) E'(E^{-1}(Q^{\alpha\beta}))}. \label{inverserate}
\end{equation}
where $q^*=|\ve{p}+\ve{p}'-\ve{q}|=|\ve{P}-\ve{q}|$. We normalized the rate to unity by the definition (\ref{inverserate}).
The quantity $Q^{\alpha\beta}$ is the unique solution of the equation $K^{\alpha\beta}=E(q)\oplus^{\alpha\beta}E(Q^{\alpha\beta})$, indicated by the modified energy constraint. $y^*$ is the value of $y=\frac{P^2+q^2-|\ve{P}-\ve{q}|^2}{2Pq}$, when the energy constraint is satisfied. The function in the denominator is $g^{\alpha\beta}(x,y)=\partial_y x\oplus^{\alpha\beta}y$. We use a short-hand notation for the step function: $\Theta(|y|\leq 1)=\theta(1+y)\theta(1-y)$ with $\theta(.)$ being the Heaviside step-function. $x$ is defined by $\ve{p}\cdot\ve{p}'=:pp'x$. The definition Eq. (\ref{inverserate}) makes $\mathcal{Z}^{\alpha\beta}$ to the surface area of the constraint surface $\mathcal{C}^{\alpha\beta}$.\\ \indent
Now we specify the energy addition rule. As it was emphasized in \cite{biro_comp_rule}, we use a rule which gives the simple addition for low energies. Therefore the simplest choice is 
\begin{equation} \label{addrule}
E\oplus^{\alpha\beta} E'=E+E'+A^{\alpha\beta}EE'.
\end{equation}
The quantities $A^{\alpha\beta}$ may depend on the phase-space or other dynamical details. $A^{\alpha\beta}\equiv 0$ means conventional addition. It is worth to mention that the rule (\ref{addrule}) is also easy tractable, namely its inverse function can be constructed analytically. The characteristic scales are the total energy per particle $\avr{E} =\sum_\alpha\int\!\!\mathrm{d}^3\ve{p}E(p)f^\alpha(E(p))$ and $1/A^{\alpha\beta}$. \\ \indent
Let us briefly summarize the numerical method we used to solve (\ref{kinequ3}). Since our model is homogeneous in space, we do not have to deal with the propagation path of the particles. A cascade method, following the evolution of the system collision-by-collision, is also satisfactory. The usual method, i.e. considering the collision in the center of the mass system, is not convenient, because the modified energy composition rule can be implemented problematically only. The problem manifests in the $\frac{1}{P}$ asymptotic of (\ref{inverserate}). We rather improve the method described in \cite{biro2}. Its key element is the distribution defined on the constraint surface $\mathcal{C}^{\alpha\beta}$, parametrized by the energies of the outgoing particles. Because of the simple form of the rate function only the kinematics restricts the collisions. The rate defined in (\ref{inverserate}) implies a uniform distribution on $\mathcal{C}^{\alpha\beta}(P,K)$, which has a non-trivial density function in the energy variable. Let us denote this function by $\rho^{\alpha\beta}(\epsilon,P,K,x)$. Viewing (\ref{inverserate}), as the normalization condition for $\rho^{\alpha\beta}$, and using (\ref{addrule}) with the notation $\epsilon=E(q)$, the density reads as
\begin{equation} \label{distronC}
\rho^{\alpha\beta}(\epsilon,P,K,x) =\frac{\Theta(|y^*(\epsilon,P,K,x)|\leq 1)}{\mathcal{Z}^{\alpha\beta}(P,K,x)}\frac{1}{1+A^{\alpha\beta}\epsilon}.
\end{equation}
The energy sampling for $\epsilon$ on the constraint surface was implemented using the rejection method. We simply select the energy for one outgoing particle randomly according to $\rho^{\alpha\beta}$, the quasi-energy conservation provides the other one. \\ \indent
In the previous section we argued that a detailed balance state may not exist for two components. Nevertheless the question, what will happen long time after the initial state was prepared, should be answered. Our cascade simulation provides the digitalized version of $f^\alpha(E)$ collision-by-collision. The moments of the density function are also excellent tools for the qualitative analysis, we present $\avr{E}$ in several examples. It is also useful to follow the entropy-like quantity $~S_E:=-\sum_\alpha\int\!\!\mathrm{d}EF^\alpha(E)\ln F^\alpha(E)$, where we used the density normed as $~\int\!\!\mathrm{d}^3\ve{p}f^\alpha(E(p))=:\int\!\!\mathrm{d}EF^\alpha(E)=1$.

\newpage
\subsection{Scaling solutions}
\begin{figure}
  \centering
  \includegraphics[width=.6\linewidth]{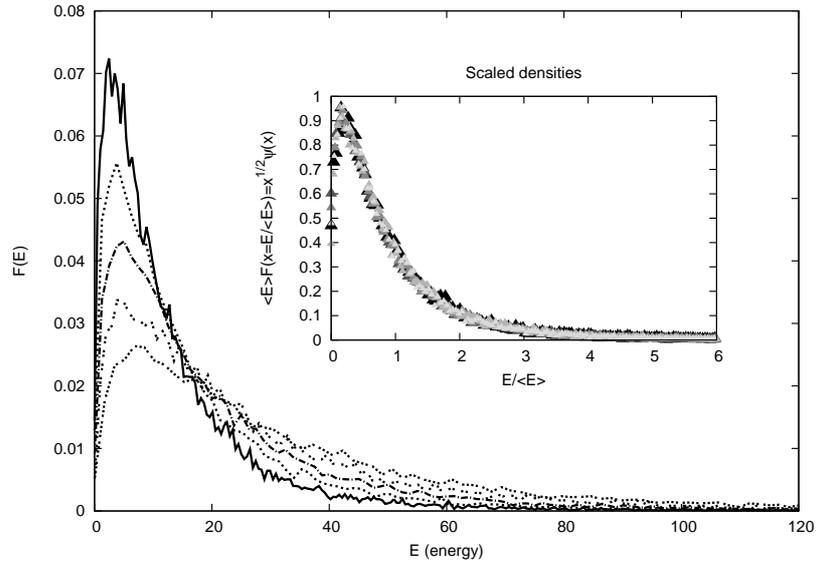}
  \caption{The density function $F(E,t)$ in different collision times, $a_0=0$, $a_1=1$. The scaling behaviour is apparent when one uses the relation (\ref{numscalingSOL}), as can be seen on the inner graph.}
  \label{fig:scale_dens_func}
\end{figure}

\begin{figure}
  \centering
  \includegraphics[width=.6\linewidth]{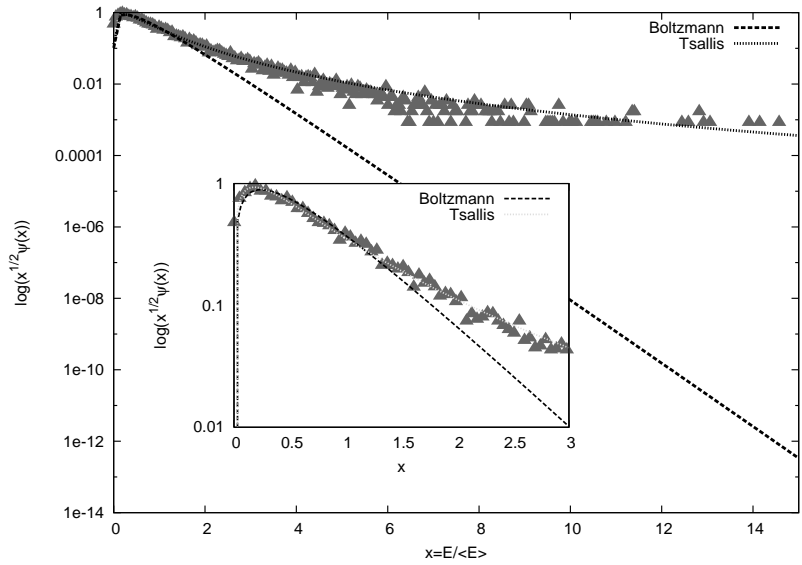}
  \caption{The asymptotic behaviour of $F(E,t)$ fitted by a Tsallis-like function $\sim\sqrt{x}(1+Ax)^{-\frac{B}{A}}$. The Boltzmann-Gibbs behaviour ($\sim\sqrt{x}e^{-Bx}$) is also indicated.}
  \label{fig:dens_func_fit}
\end{figure}

\begin{figure}
  \centering
  \includegraphics[width=0.6\linewidth]{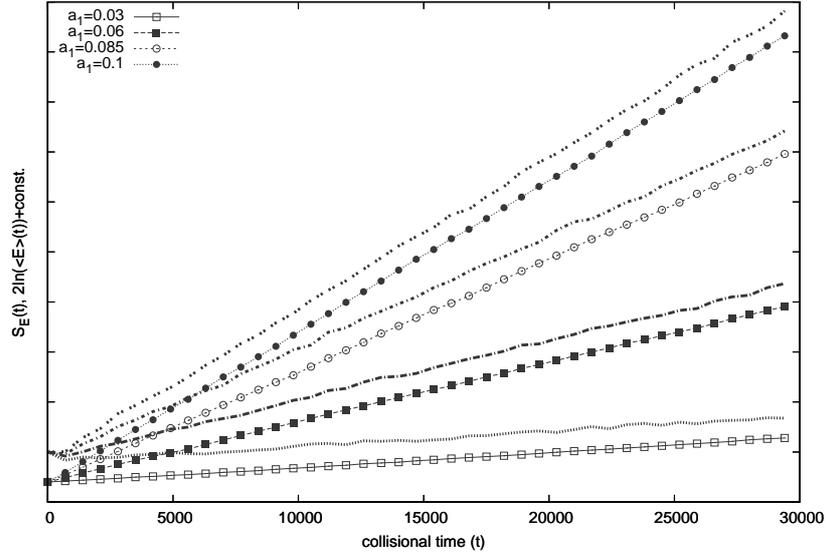}
  \caption{Parametrized equation of states: $S_E(t)$, $\ln\avr{E}(t)$ for $a_0=0$, varying $a_1$. The entropy curves run in line with the logarithm of the total energy per particle after $t_{\mathrm{trans.}}$.}
  \label{fig:entropy}
\end{figure}

\indent We analyse two cases as summarized in the following table (with $a_0$, $a_1$, $\Lambda_0 >0$ constants): 
\begin{table}[h]\centering
\begin{tabular}{c|c|c}
 & $I.)$ ($a_0$, $a_1$) & $II.)$ ($a_0$, $a_1$, $\Lambda_0$) \\
 \hline\hline
 $A^{AA}=A^{BB}$ & $\frac{a_0}{\avr{E}}$ & $\frac{a_0}{\avr{E}}$ \\
 \hline
 $A^{AB}=A^{BA}$ & $\frac{a_1}{\avr{E}}$ & $\begin{array}{lcl} \frac{a_0}{\avr{E}}, & \textnormal{if} & E_1+E_2<\Lambda_0\avr{E} \\ \frac{a_1}{\avr{E}}, &\textnormal{if} & E_1+E_2>\Lambda_0\avr{E}  \end{array}$  \\
 \hline
$\begin{array}{c}\textnormal{long-time}\\ \textnormal{behaviour} \end{array}$ & $\begin{array}{c}\textnormal{growing}\\ \avr{E} \end{array}$ & $\begin{array}{c}\textnormal{growing or lowering}\\ \avr{E} \end{array}$
\end{tabular}
\caption{The two investigated cases with scaled modification parameters. $a_0$, $a_1$ and $\Lambda_0$ are positive.}
\label{tab:tab1}
\end{table}
The numerical investigation shows that in case $I.)$ the total energy per particle is always growing for long times ($a_0$, $a_1$ are positive constants), cf. Fig. (\ref{fig:entropy}). Depending on the value of $\Lambda_0$, either growing or lowering of the energy per particle can occur in the case $II.)$ ($\Lambda_0$ is a positive constant), cf. Fig. (\ref{fig:en_saturation}). \\
Each one of the modifications $I.)$ and $II.)$ has an interesting feature. Both result in scaling density functions, insensitive to the initial conditions:
\begin{equation} \label{numscalingSOL}
f^{A,B}(E,t\rightarrow \infty) \sim \avr{E}^{-\frac{3}{2}}\psi(E/\avr{E}), 
\end{equation}
in other words the long-time evolution is defined by a one-parameter family of density functions. The $\psi$ shape-function is time-independent, the only time-dependence occurs due to $\avr{E}$. \\
In fact, the scaling behaviour (\ref{numscalingSOL}) can be observed after a transient time $t>t_\mathrm{trans.}$. $t_\mathrm{trans.}$ has the same order of magnitude as the relaxation time to the detailed balance state in one-component systems ($\mathcal{O}(10)$ collision per particle, in collision time). Since there is no parameter which could distinguish among the two components $A$ and $B$, we expect the same asymptotic density function, if a steady state evolves. We used two distinct states to prepare the initial conditions, namely a "thermal" one (Boltzmannian density) and the "two fireball" (one half of the particles moving into an assigned direction while the other half into the opposite direction). We did not experience any differences about the long-time behaviour for the various initial densities. Using the relation (\ref{numscalingSOL}) one can scale the densities belonging to different collisional times onto each other. However the scaling is conspicuous on Fig. (\ref{fig:scale_dens_func}), it has another apparent feature, namely that the entropy is related to the total energy per particle as $S_E \sim \ln\avr{E}$, cf. Fig. (\ref{fig:entropy}). \\ \indent
It turns out that the scaling function $\psi(x)$ has power-law tail rather than an exponential one (when $x\rightarrow\infty$). The Tsallis-density function fits it quite well: ${~\psi(x) \sim \sqrt{x}(1+Ax)^{-\frac{B}{A}}}$. This function is the detailed balance solution of the one-component MBE. An in-depth analysis would be needed to derive the fit parameters $A$ and $B$ starting from the given modification parameters $a_0$, $a_1$ and $\Lambda_0$, or reveal the numerically hardly visible bias from the Tsallis fitting function (Fig. (\ref{fig:dens_func_fit})).\\ \indent
The existence of the solution (\ref{numscalingSOL}) can be proven by using the algebraic identity $~E\oplus_{\avr{E}}^{\alpha\beta}E'= \avr{E}\left(\frac{E}{\avr{E}}\oplus_1^{\alpha\beta}\frac{E'}{\avr{E}}\right)$ valid for Eq. (\ref{addrule}) if $A^{\alpha\beta}=\frac{a^{\alpha\beta}}{\avr{E}}$, as can be found in Appendix \ref{app2} in details. \\
Numerical experiences indicate that such a family of density functions is asymptotically stable under time evolution. Due to (\ref{numscalingSOL}), $~\avr{E}(t) = \avr{E}_0 e^{\gamma (t-t_0)}$ holds for the collision time-evolution. We discuss the connection between collision time and laboratory time in Appendix \ref{app3}. We note here that such scaling solutions occur in the context of non-elastic Boltzmann equation (NEBE), referred as homogeneous cooling (heating) states \cite{ernst,bobylev1,bobylev2}. Both in NEBE and MMBE models the total kinetic energy is not a conserved quantity -- the modelled systems are open.
\subsection{Heating, cooling and saturation}
\begin{figure}
  \centering
  \includegraphics[width=0.6\linewidth]{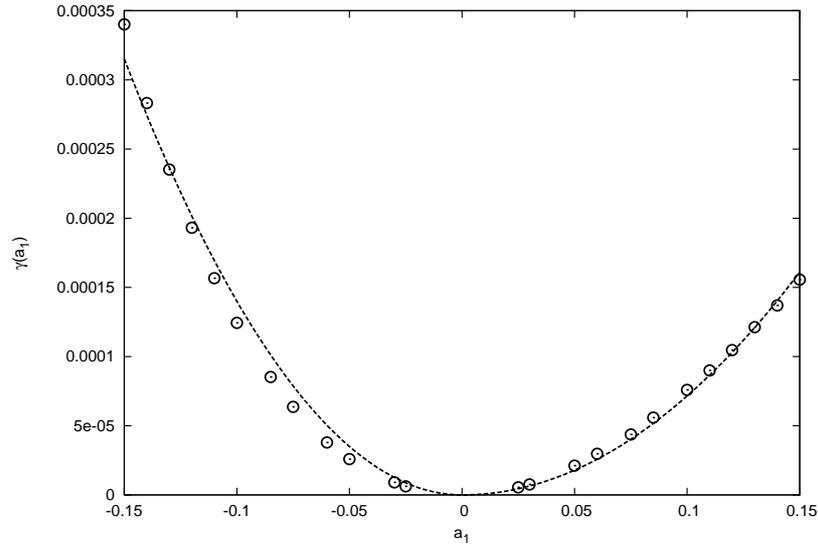}
  \caption{The exponent $\gamma$ for $a_0=0$ and different values of $a_1$. It approaches zero as $\gamma \propto a_1^2$.}
  \label{fig:gamma_vs_a}
\end{figure}

Now we turn to the interpretation of the scaling solutions (\ref{numscalingSOL}), which apparently rule the long-time behaviour. This solution is characterized by the exponent $\gamma$ and the initial condition, say $\avr{E}_0$. The values of $a_0$, $a_1$ affect the shape function $\psi$ and the value of $\gamma$. Depending on the modification parameters, it disappears smoothly: $\gamma \simeq C^\pm(a_1-a_0)^2$ when $~a_1-a_0 \rightarrow 0^\pm$, as it is demonstrated in Fig. (\ref{fig:gamma_vs_a}). It is easy to see that for the $\kappa$-th moments of $f(E,t>t_\mathrm{trans.})$ in the scaling regime $~\avr{E^\kappa} \sim \avr{E}^\kappa$ holds. Since the system is homogeneous even in the momentum-space, the macroscopic behaviour can be described by these moments. As the time evolution becomes trivial, the following equation of state stabilizes (Fig. (\ref{fig:entropy})):
\begin{eqnarray} \label{EoS}
S_E &=& 2C_1\ln\avr{E} +C_2, \textnormal{ with}\\
C_1=\left(\int_0^\infty\!\!\mathrm{d}x\sqrt{x}\psi(x)\right) \ln\left(\int_0^\infty\!\!\mathrm{d}x x^\frac{3}{2}\psi(x)\right) &\textnormal{and}& C_2=-\int_0^\infty\!\!\mathrm{d}x \left(\sqrt{x}\psi(x)\ln(\sqrt{x}\psi(x))\right), \nonumber
\end{eqnarray}
where $C_1$ and $C_2$ are time- and phase-space-independent quantities. \\
\textit{That is, the system behaves like an ideal gas, which is heating up or cooling down depending on the sign of $\gamma$.}
\subsection{Saturation}
\begin{figure}
  \centering
  \includegraphics[width=0.6\linewidth]{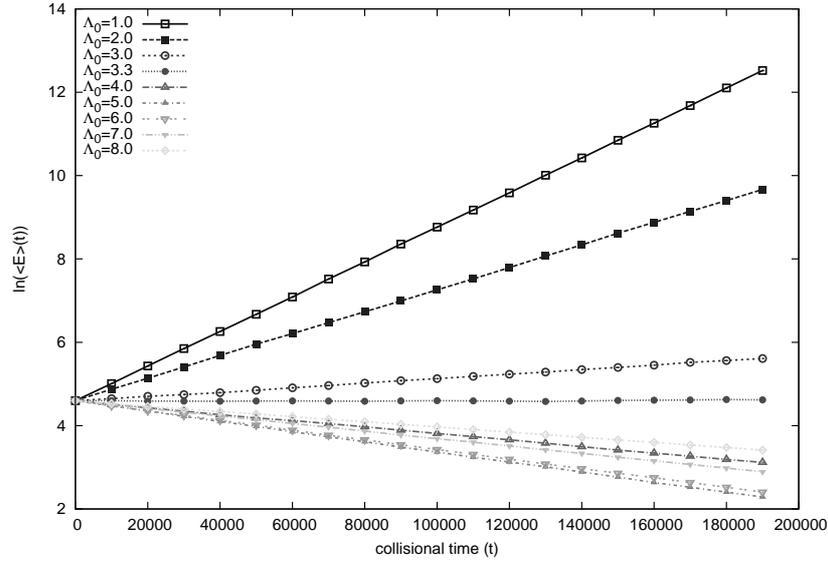}
  \caption{The running of $\ln\avr{E}(t)$ for various value of $\Lambda_0$. The total energy per particle can either be growing or lowering with time. With the fine-tuning of $\Lambda_0$ saturation occurs.}
  \label{fig:en_saturation}
\end{figure}

\begin{figure}
  \centering
  \subfloat[]{
	  \includegraphics[width=0.49\linewidth]{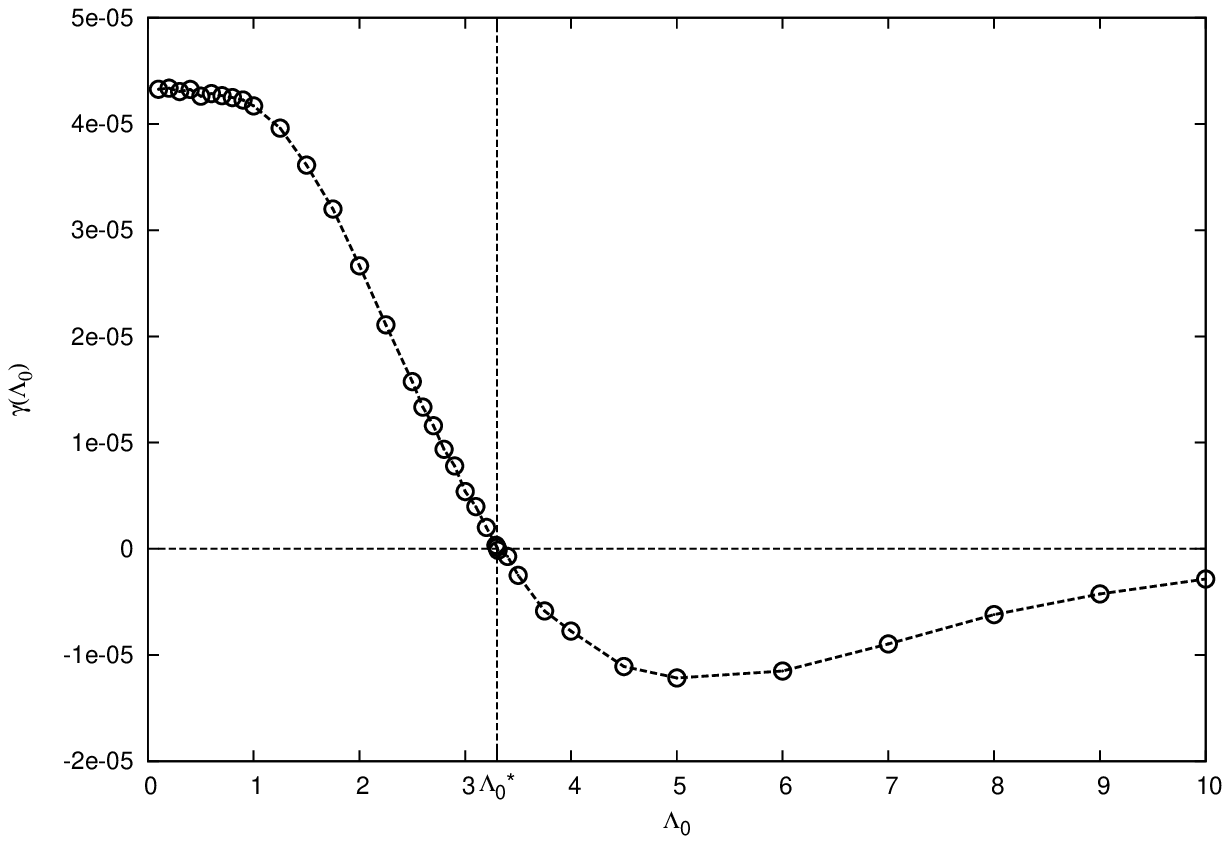}
	  \label{fig:gamma_vs_Lambda}
	  }
   \subfloat[]{
	\includegraphics[width=.49\linewidth]{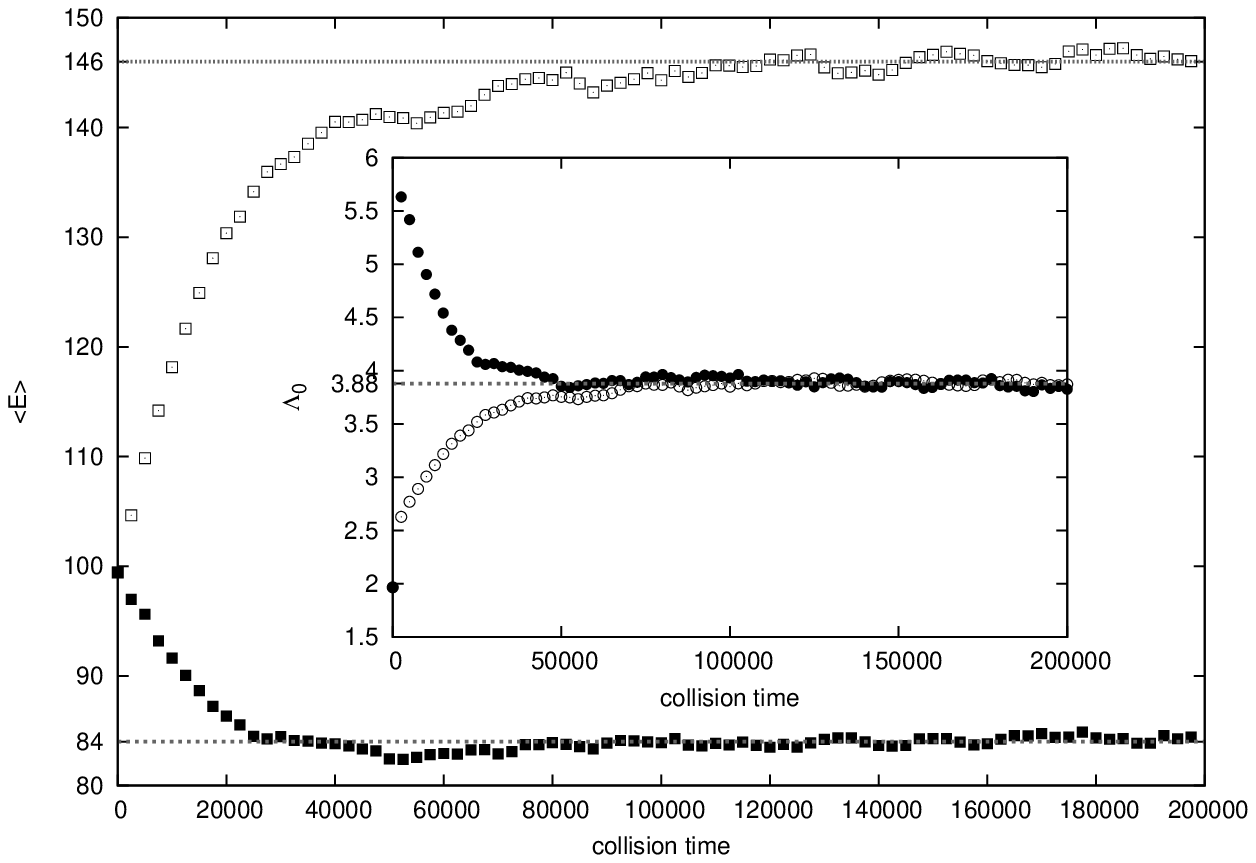}
	\label{fig:satur_feedback}
	}
  \caption{\textit{(a)}: The exponent $\gamma$ as a function of $\Lambda_0$. At its zero, the total energy per particle stands still. \textit{(b)}: Feedback of the lowering or growing of $\Lambda_0$ as $\avr{E}$ varies. We used a simple smoothed step-function to mimic the feedback effect $\dot{\Lambda}_0 \propto \dot{\avr{E}}$. Although the system is in a steady state for a long time -- compared to the saturation to that state --, the energy per particle starts to grow slowly, because $\dot{\avr{E}}$ -- and therefore $\Lambda_0$ -- has large fluctuations. ($a_0=0.11$, $a_{10}=0.21$, $\Lambda_0^*=3.88$.)}
\end{figure}

The main difference between the different choices of the modification (cf. Table \ref{tab:tab1}) is the following. In the case $I.)$, the section of the constraint surfaces $\mathcal{C}^{AA}(\avr{E})$ and $\mathcal{C}^{AB}(\avr{E})$ is empty (or at least its surface measure is zero), while ${~\mathcal{C}^{AA}(\avr{E}) \rightarrow \mathcal{C}_0 \leftarrow \mathcal{C}^{AB}(\avr{E})}$ for $\avr{E} \rightarrow \infty$. Here $\mathcal{C}_0$ is the constraint surface in the unmodified case $a_0=0$, $a_1=0$. In 3-dimension with $E \propto p^2$ dispersion $\mathcal{C}_0$ is a sphere. \\ \indent
In the case $II.)$ with interaction threshold, there is a non-zero section of $\mathcal{C}^{AA}(\avr{E})$ and $\mathcal{C}^{AB}(\avr{E})$ for any time. While in case $I.)$ the system reaches a steady state without detailed balance, it turned out by the numerical investigation that in case $II.)$ an equilibrium state develops if $\Lambda_0$ is fine-tuned. Since the energy cut-off was also scaled out with the average energy $\avr{E}$, the scaling behaviour (\ref{numscalingSOL}) prevails. We depicted the running of $\avr{E}$ and $\gamma$ for various values of $\Lambda_0$ on Fig. (\ref{fig:en_saturation}). As it can be seen on Fig. (\ref{fig:gamma_vs_Lambda}), $\gamma(\Lambda_0)$ has a zero, in the present example ($a_0=0.15$, $a_{10}=0.25$) at $\Lambda_0^* \approx 3.3$. A qualitative explanation of this behaviour can be given if one takes notice of the fixed shape of the density function in the scaled variable $E/\avr{E}$. That is why the probability of such collisions, where the total kinetic energy grows (or decreases), is also constant in the scaling regime, being proportional to the integral of $\psi(E/\avr{E})$ on a definite domain. Therefore the probability depends on the modification parameters only, as all the quantities in the scaling regime do. $\Lambda_0$ prescribes that how much the two distinct type of collisions featured by $a_0$ and $a_1$, respectively contribute to the probability of energy growing (decreasing) in the corresponding energy ranges $E_1+E_2 < \Lambda_0\avr{E}$ and $E_1+E_2 > \Lambda_0\avr{E}$. If the kinetic energy domain, which is responsible for the lowering effect, is large enough, then the energy change can be compensated statistically and the system equilibrates. \\ \indent
Multicomponent kinetic equations derived on first-principle basis show thermalization and develop equilibrium state for large times \cite{cassing2}. In the case $II.)$ it is possible to make the fixed point in the $\Gamma_0-\gamma$ phase-space attractive due to dynamical feed-back. Increasing $\Lambda_0$ when the energy is growing and lowering it if $\avr{E}$ is decreasing makes $\gamma(\Lambda_0^*)=0$ to be a stable fixed point of the time evolution. If the system relaxes to a scaling state with a different $\gamma$ fast enough when $\Lambda_0$ changes, then $\gamma(\Lambda_0)$ is indeed the allowed phase-space in the scaling regime. The result of this very simple feed-back, $~\dot{\Lambda}_0 \propto \dot{\avr{E}}$ can be seen in Fig. (\ref{fig:satur_feedback}) from numerical simulation. $\avr{E}$ and $\Lambda_0$ tend to a constant value, as it is expected. 

\section{Conclusion}
\indent We put an emphasis in this paper on investigating whether an MMBE system tends to an equilibrium state or not. It is a non-trivial question, even with the conceptually simplest modification of the two-particle constraints (modifying the energy addition rule). As far as we know, though the modification of the BE due to the constraints is well discussed in the literature through several examples, there are no studies concerning modified, multicomponent systems. The problem arises for all studied examples (either energetic or entropic reasoning for the modification of the collision integral) \cite{biro3,kaniadakis,biro4,rafelski}. The detailed balance state in multicomponent case is generally lacking because different conditions are to be satisfied for each piece of the collision integral to vanish. \\ \indent 
Our conclusion is that dealing with such kind of MMBE, equilibration is not guaranteed in general. The same quantity which is conserved in the one-component system with a non-additive energy composition rule, in the multicomponent case describes an open system. In order to achieve a stationary state, one has to go beyond the simple kinetic treatment, and has to feedback the dynamics of the energy non-additivity to the MMBE. It is conceivable that for a satisfactory description one has to return to the microscopic description of the off-mass-shell effects. 

\subsection*{Acknowledgements}
M.H. would like to thank to Antal Jakov\'ac, P\'eter V\'an, K\'aroly \"Urm\"ossy and P\'eter Mati for all the discussions, motivations and help during the preparation of his very first paper. \\
M.H. was supported financially by New Sz\'echenyi Plan (Project no. T\'AMOP-4.2.2.B-10/1--2010-0009). This work was supported by the Hungarian Research Fund (OTKA) under contract No. K104260.

\appendix
\section{Approximating $\llangle \mathcal{I}_\gamma \rrangle$}\label{app1}
In the following we discuss the averaging and the approximations we used calculating $\mathcal{I}_\gamma$ and $\mathcal{I}$, which results the formula (\ref{averaging}) at the end. Using compact notations, the 9-fold collision integral reads:
\begin{eqnarray}
\mathcal{I}_\gamma &=& \int_{234}\!\!\int\!\!\mathrm{d}\omega\rho_\gamma(\omega -E_2)\delta(E_1+\omega -E_3-E_4)\delta^{(3)}({\sum_i}'\ve{p}_i)w_{1234}(f_3f_4-f_1f_2)=: \int_{234}\!\!\rho_\gamma(E_3+E_4-E_1-E_2)\mathcal{K}_{1234}
\end{eqnarray}
We perform now the averaging according to $\gamma$: 
\begin{eqnarray}
\mathcal{I} &=& \int\limits_0^\infty\!\!\mathrm{d}\gamma g(\gamma)\mathcal{I}_\gamma =\int_{234}\!\!\mathcal{K}_{1234} \int_0^\infty\!\!\mathrm{d}\gamma g(\gamma)\rho_\gamma(E_3+E_4-E_1-E_2) =:\int_{234}\!\!\mathcal{K}_{1234} \llangle\rho_\gamma\rrangle(E_2-\Omega) \approx \label{1stline}\\
& \approx & \int_{234}\!\!\left\{\left.\mathcal{K}_{1234}\right|_{E_2=\Omega+\Delta} +\partial_{E_2}\left.\mathcal{K}_{1234}\right|_{E_2=\Omega+\Delta}(E_2-\Omega-\Delta)+ \mathcal{O}((E_2-\Omega-\Delta)^2) \right\}\llangle\rho_\gamma\rrangle(E_2-\Omega) = \label{2ndline}\\
&=& \int_{234}\!\!\delta({\sum_i}'E_i-\Delta)\mathcal{K}_{1234} + \left\{ \,\,\mathrm{terms\,proportional\,to\,higher\,moments\,of}\,\gamma\,\, \right\}. \label{3rdline}
\end{eqnarray}
In the first line (Eq. (\ref{1stline})) the interchangeability of the integrations respect to $\gamma$ and to the phase-space coordinates is assumed. Let us suppose the function $\llangle\rho_\gamma\rrangle$ to be a density function with the properties $\int_0^\infty\!\!\mathrm{d}\omega \llangle\rho_\gamma\rrangle(\omega)=1$ and $\int_0^\infty\!\!\mathrm{d}\omega \llangle\rho_\gamma\rrangle(\omega)\omega=\Delta$. Furthermore, let its higher moments be negligible compared to $\Delta\gg \int_0^\infty\!\!\mathrm{d}\omega \llangle\rho_\gamma\rrangle(\omega)\omega^n$. These imply that $\llangle\rho_\gamma\rrangle$ is "peaky" around $\Delta$. Therefore in Eq. (\ref{2ndline}) we expand the rest of the integral kernel in $E_2-\Omega$ around $\Delta$. After integrating respect to $E_2$, the first term gives the kernel $\mathcal{K}$ evaluated in $E_2=\Omega+\Delta$, the second vanishes. The remaining terms are at least $\mathcal{O}(\Delta^2)$. In the last line (Eq. (\ref{3rdline})) we reimplemented the effect of integrating respect to $E_2$ as resulted by a modified constraint for the kinetic energies.

\section{Scaling symmetry}\label{app2}
The aim of this section is to show the existence of a special family of solutions of Eq. (\ref{kinequ3}) with a scaling property and the joint pattern of the time evolution. For this purpose, we assume that the function $f^\alpha(E,t_0)$ is a solution of Eq. (\ref{kinequ3}). Then using the properties of the collision integral we prove that $\avr{E}^{3/2}f^\alpha(\avr{E}E,t_0)$ is also a solution with $\avr{E}=\avr{E}_0e^{\gamma(t-t_0)}$ for arbitrary $t$ with the time-independent constant $\gamma$. We do not investigate in this section the stability of the scaling solution, only mention that it is indicated by our numerical simulations. \\
We consider the MMBE (\ref{kinequ3}) with energy variables:
\begin{eqnarray} \label{kinequ4}
\partial_{t} f^\alpha(E,t) &=& \sum_\beta N^{\alpha\beta}\int_0^1\!\!\mathrm{d}x \int_0^1\!\!\mathrm{d}y \int_0^\infty\!\!\mathrm{d}E' \int_0^\infty\!\!\mathrm{d}\epsilon \frac{\sqrt{E'\epsilon}}{\mathcal{Z}_{\lambda}^{\alpha\beta}(E,E',x)}\delta(K^{\alpha\beta}_{\lambda}(E,E')-\epsilon\oplus^{\alpha\beta}_{\lambda}E^*(E,E',\epsilon,x,y)) \times \nonumber \\
& & \times (f^\alpha(\epsilon,t)f^\beta(E^*,t)-f^\alpha(E,t) f^\beta(E',t)) =: \sum_\beta\mathcal{I}_{\lambda}^{\alpha\beta}[f^{\alpha,\beta}(\,.\,,t)](E).
\end{eqnarray}
Here $\lambda=\lambda(t)$ having the dimension of energy, $N^{\alpha\beta}$ is a constant (from the spherical integration and the dispersion relation). Finally 
\begin{eqnarray}
E^*=E(|\ve{P}-\ve{q}|)=\frac{1}{2m}(P^2+2m\epsilon-2P\sqrt{2m\epsilon}y), & \,\,\,\, & P=\sqrt{2m}\sqrt{E+E'+2EE'x}. \nonumber
\end{eqnarray}
Using the easily verifiable identity $\mathcal{Z}^{\alpha\beta}_\lambda(E,E',x)= \sqrt{\lambda}\mathcal{Z}^{\alpha\beta}_1\left(\frac{E}{\lambda},\frac{E'}{\lambda},x\right)$, then changing the variables of integration, the following holds ($\lambda'=\lambda(t')$):
\begin{equation} \label{scalingI}
\left(\frac{\lambda}{\lambda'}\right)^{\frac{3}{2}}\mathcal{I}^{\alpha\beta}_{\lambda}[f^{\alpha,\beta}(\, .\,,t)](E) = \mathcal{I}^{\alpha\beta}_{\lambda'}\left[ \left(\frac{\lambda}{\lambda'}\right)^{\frac{3}{2}}f^{\alpha,\beta}\left(\frac{\lambda}{\lambda'}(\, .\,),t\right) \right]\left(\frac{\lambda'}{\lambda}E \right).
\end{equation}
One takes then Eq. (\ref{kinequ4}) and multiplies it by the factor $(\frac{\lambda}{\lambda'})^{\frac{3}{2}}$. Using (\ref{scalingI}) one arrives at
\begin{equation} \label{scalingSol}
\left(\frac{\lambda}{\lambda'}\right)^{\frac{3}{2}}\partial_{t}f^{\alpha}(E,t) = \sum_\beta \mathcal{I}^{\alpha\beta}_{\lambda'}\left[ \left(\frac{\lambda}{\lambda'}\right)^{\frac{3}{2}} f^{\alpha,\beta}\left(\frac{\lambda}{\lambda'}(\, .\,),t\right) \right]\left(\frac{\lambda'}{\lambda}E \right) =: \sum_\beta \mathcal{I}^{\alpha\beta}_{\lambda'}[ f^{\alpha,\beta}((\, .\,),t') ]\left(\frac{\lambda'}{\lambda}E \right) = \partial_{t'}f^{\alpha}(\lambda'\lambda^{-1}E,t').
\end{equation}
One realizes that the first equality in Eq. (\ref{scalingSol}) is the MMBE (\ref{kinequ4}) with rescaled energy-argument. Eq. (\ref{scalingSol}) suggests the following scaling relation to hold:
\begin{equation} \label{scalingDF0}
f^\alpha(E,t')=\left(\frac{\lambda(t)}{\lambda(t')} \right)^{\frac{3}{2}} f^\alpha\left(\frac{\lambda(t)}{\lambda(t')} E,t\right).
\end{equation}  
In other words, the rescaling of the energy variable means a finite step of time-evolution. We have to check the conditions of the equivalence of Eq. (\ref{scalingDF0}) and Eq. (\ref{scalingSol}). Let us differentiate Eq. (\ref{scalingDF0}) respect to $t'$:
\begin{eqnarray} \label{checkscaling}
\partial_{t'} f^\alpha\left(\frac{\lambda(t)}{\lambda(t')}E,t'\right) &=& f^\alpha(E,t')\partial_{t'}\left[\left(\frac{\lambda(t)}{\lambda(t')}\right)^{\frac{3}{2}}\right] +\left(\frac{\lambda(t)}{\lambda(t')}\right)^{\frac{3}{2}} \frac{\mathrm{d}t}{\mathrm{d}t'}\partial_t f^\alpha(E,t) = \nonumber \\
&=& \frac{3}{2}\left(\frac{\lambda(t)}{\lambda(t')}\right)^{\frac{3}{2}}\left(\frac{\partial_t\lambda(t)}{\lambda(t)}\frac{\mathrm{d}t}{\mathrm{d}t'}-\frac{\partial_{t'}\lambda(t')}{\lambda(t')}\right)f^\alpha(E,t') +\left(\frac{\lambda(t)}{\lambda(t')}\right)^{\frac{3}{2}} \frac{\mathrm{d}t}{\mathrm{d}t'}\partial_t f^\alpha(E,t) = \nonumber \\
&=& \left(\frac{\lambda(t)}{\lambda(t')}\right)^{\frac{3}{2}} \partial_t f^\alpha(E,t).
\end{eqnarray}
The last equality in Eq. (\ref{checkscaling}) holds when $\frac{\mathrm{d}t}{\mathrm{d}t'} \equiv 1$. and $\partial_t\lambda/\lambda \equiv \mathrm{const.}$ In summing up, the relation in Eq. (\ref{scalingDF0}) holds if $t'=t_0+t$ and $\lambda(t')=\lambda_0e^{\gamma(t'-t_0)}$, with arbitrary constants $t_0$, $\lambda_0$ and $\gamma$.\\
It is easy to establish the connection between $\lambda$ and the energy density $\avr{E}$ for scaling solutions. After changing the variable in the integrand one gets
$$\lambda(t) \sim \avr{E}(t)= \left(\sum_\alpha 2\pi(2m)^\frac{3}{2}\int\!\!\mathrm{d}x\sqrt{x}x f^\alpha(x=E\lambda)\right)\lambda,$$
so $\lambda$ is simply proportional to $\avr{E}$. The expression in the parenthesis is constant due to Eq. (\ref{scalingDF0}), choosing $\lambda(t')=1$. \\
In conclusion:
\begin{eqnarray} \label{scalingDF}
f^\alpha(E,t) =\frac{1}{\avr{E}^{3/2}(t)}\psi^\alpha\left(\frac{E}{\avr{E}(t)} \right), & & \avr{E}(t)=\avr{E}_0 e^{\gamma(t-t_0)},\end{eqnarray}
where $t_0$ is used as a reference point: $\avr{E}_0=\avr{E}(t_0)$ with the definition $~\psi^\alpha(x)=\avr{E}_0^{3/2}f^\alpha(\avr{E}_0 x,t_0)$.

\section{Time evolution in laboratory time}\label{app3}
\indent In this study we used the collision frequency to follow the evolution of the density function of the system and to analyse the long-time behaviour of the total energy per particle $\avr{E}$. Still, the relation of the collision counter $t$ to the laboratory time $t_\mathrm{lab}$ needs clarification. The quantity which gives exactly the probability of a (given kind of) collision in a unit \textit{laboratory} time is nothing but $w^{\alpha\beta}=(\mathcal{Z}^{\alpha\beta}_\lambda)^{-1}$. So the expected number of collisions in a unit time in a given state of the system is given by
\begin{eqnarray}
\frac{\mathrm{d}t}{\mathrm{d}t_\mathrm{lab}} &=& \sum_{\alpha\beta}\int\!\!\mathrm{d}^3\ve{p}\int\!\!\mathrm{d}^3\ve{p}'w^{\alpha\beta}(\ve{p},\ve{p}')f^\alpha(E(p))f^\beta(E(p')) \sim \lambda^{-\frac{1}{2}}(t) \int_0^\infty\!\!\mathrm{d}\omega\sqrt{\omega} \int_0^\infty\!\!\mathrm{d}\eta\sqrt{\eta} \int_0^1\!\!\mathrm{d}x \frac{\psi(\omega)\psi(\eta)}{\mathcal{Z}_1^{\alpha\beta}(\omega,\eta,x)},
\end{eqnarray}
for scaling solutions described by (\ref{numscalingSOL}). This proportionality leads us to the following separable ODE:
\begin{equation}
\frac{\mathrm{d}t}{\mathrm{d}t_\mathrm{lab}}=\Gamma \avr{E}^{-\frac{1}{2}}(t)=\Gamma\avr{E}_0^{-\frac{1}{2}}e^{-\frac{\gamma}{2}t}
\end{equation}
Here $\Gamma$ is constant in time, it is however a fairly difficult task to determine it. One has to solve an integro-differential equation for the shape function $\psi(x)$ to get $\Gamma$. But if its value is assumed to be known, the collision time expressed by the laboratory time (with the initial condition $t_\mathrm{lab}(t=0)=0$) reads as:
\begin{equation}\label{colltime_labtime}
t=\frac{2}{\gamma}\ln\left(1+\frac{\Gamma\gamma}{2\avr{E}_0^{1/2}}t_\mathrm{lab}\right).
\end{equation}
Using the expression (\ref{colltime_labtime}) the energy per particle in laboratory time becomes
\begin{equation}\label{en_density_labtime}
\avr{E}(t_\mathrm{lab})= \avr{E}_0\left(1+\frac{\Gamma\gamma}{2\avr{E}_0^{1/2}}t_\mathrm{lab}\right)^2.
\end{equation}
For $\gamma>0$ the energy per particle is growing limitless like $\sim t_\mathrm{lab}^2$. Let us remark that for $t_{\mathrm{lab}} \gg \frac{2\avr{E}_0^{1/2}}{\Gamma\gamma}$, $\avr{E}(t_{\mathrm{lab}}) \sim (\Gamma\gamma t_{\mathrm{lab}})^2$ is $\avr{E}_0$-independent. \\
For $\gamma<0$, $\avr{E}$ reaches zero in a finite time in the laboratory system: $t_\mathrm{lab}^*=\frac{2\avr{E}_0^{1/2}}{\Gamma|\gamma|}$. It is interesting, that the number of collisions tends to infinity as $t_\mathrm{lab} \rightarrow t_\mathrm{lab}^*$ in this case. This indicates the presence of the so-called inelastic collapse in the context of NEBE \cite{inelcoll}.

\end{document}